**A Dynamical Systems Approach to Bots and Online Political Communication**

Beril Bulat

University of California, Davis

bbulat@ucdavis.edu

Martin Hilbert

University of California, Davis

hilbert@ucdavis.edu

## Abstract

Bots have become increasingly prevalent in the digital sphere and have taken up a proactive role in shaping democratic processes. While previous studies have focused on their influence at the individual level, their potential macro-level impact on communication dynamics is still little understood. This study adopts an information theoretic approach from dynamical systems theory to examine the role of political bots shaping the dynamics of an online political discussion on Twitter. We quantify the components of this dynamic process in terms of its complexity, predictability, and its entropy rate, or the remaining uncertainty. Our findings suggest that bot activity is associated with increased complexity and uncertainty in the structural dynamics of online political communication. This work serves as a showcase for the use of information-theoretic measures from dynamical systems theory in modeling human-bot dynamics as a computational process that unfolds over time.

## 1. Introduction

The rise of bots in the digital age has transformed the online ecosystem in profound ways. Among these, political bots have attracted particular attention due to their potential impact on public opinion and democratic processes (Woolley & Howard, 2016). Political bots are automated agents specifically tasked with public opinion manipulation. They are the epitome of the larger collection of so-called persuasive tech (Fogg, 2002; C. Yan et al., 2012). A multitude of studies so far provided evidence of bot propaganda during elections and campaigns around the world (Bastos & Mercea, 2019; Bessi & Ferrara, 2016; Ferrara, 2017; S. C. Woolley & Howard, 2016). Previous work suggests bots are capable of passing as humans and engaging in complex interactions with other users (Ferrara et al., 2016; Wischnewski et al., 2022). Among their manipulative strategies are spreading disinformation and amplifying polarizing content (Bradshaw et al., 2020; Broniatowski et al., 2018; Vosoughi et al., 2018). However, debates concerning the impact of their malicious activities remain to be settled (Duan et al., 2022; González-Bailón & De Domenico, 2021; Howard et al., 2018).

While majority of the previous work focuses on their impact at the individual level, important questions remain at the macro level. For instance, how do bots influence the overall communication dynamics? Bots are based on predictive algorithms that follow (machine learning) rules that algorithmically transform input to output, but does this make online political communication less uncertain (given bots' employment of predictive algorithms), or more uncertain? Do bots simplify the arising dynamic or does the communication landscape get more complex with bots? These questions are becoming increasingly important as the recent revolution in generative AI is unlikely to decrease the role of bots (Fui-Hoon Nah et al., 2023;

Shin et al., 2024). On the contrary, it can be expected that the new technological possibilities of generative AI will lead to rather complex collective dynamics.

To contribute to a better understanding of the arising dynamics, in this work, we follow a long tradition that views communication as a dynamical process (Cappella & Planalp, 1981; Ellis & Fisher, 1975; Fisher, 1970; Poole & Roth, 1989), and use the tools rooted in information theory, i.e. Claude Shannon's "mathematical theory of communication" (Kolmogorov, 1958; Shannon, 1948), to answer these questions. The merger of information theory with theoretical computer science, which in its roots goes back to Kolmogorov (1959, 1968) and Sinai (1959), allows us to analyze dynamical systems, by quantifying how much information and uncertainty is created and passed along a given process (Crutchfield et al., 2009). We adopt this methodological framework to analyze a case of online political discussion and quantify the arising dynamic of the system as a whole in terms of predictability, complexity and remaining uncertainty. We then assess the relationship between political bot involvement, and changes in the components of the process, the complexity and uncertainty levels of the communicative dynamics.

In line with related findings from editor bots in Wikipedia (Hilbert & Darmon, 2020), our findings reveal that political bots significantly impact the dynamics of online political communication, contributing both to its complexity and uncertainty. The presence of political bots is associated with a less predictable and stable online discourse, which can be misleading and may negatively affect public opinion formation around controversial topics. From a methodological perspective, applying a complex systems lens on political bots and online political communication highlights the need for further research into macro-level bot influence, and the importance of considering communication processes as complex dynamic systems.

## 2. Literature Review

### 2.1 A Brief History of Bots

Bots are automated agents that perform tasks online (Franklin & Graesser, 1997; Gorwa & Guilbeault, 2020). They can run continuously without requiring any human intervention, though their agency is limited, and they can only act within the boundaries pre-defined in their scripts (Tsvetkova et al., 2017). While they have surged in popularity and visibility in recent years, their existence goes back to the advent of computing technology (Ferrara et al., 2016; Gorwa & Guilbeault, 2020). And they have come a long way since the introduction of Weizenbaum's Eliza (1966), the first chatbot. Today, bots are integral to a wide range of online applications, such as providing virtual customer service or automating content moderation (Makhortykh et al., 2022; Schanke et al., 2021). Their evolution reflects a broader transformation in their use and overall impact, which greatly expanded with the rise of social media.

The emergence of social media platforms introduced social bots that automate content production and engagement, blurring the lines between genuine human communication and artificial interaction. Social bots can be defined as automated agents that imitate human behavior on social media (Boshmaf et al., 2011). They can serve benign purposes, like news bots that distribute news articles (Hepp, 2020; Lokot & Diakopoulos, 2016), or editor bots that oversee published content (Tsvetkova et al., 2017). Or they can have more insidious purposes, such as spam attacks (Zhang et al., 2013), identity theft (Goga et al., 2015), or public opinion manipulation (Bessi & Ferrara, 2016; Ratkiewicz et al., 2011).

A subgroup of social bots, political bots have recently garnered significant attention for their impact on online discourse (S. C. Woolley & Howard, 2016). These politically motivated agents have demonstrated their capacity to amplify specific narratives, contributing to the public opinion polarization and posing challenges to the integrity of public discourse (Bessi & Ferrara, 2016; Bradshaw et al., 2020; S. C. Woolley & Howard, 2016). While several countries have attempted to enact regulations to curb computational propaganda efforts, these attempts largely fell short due to challenges with bot identification and overreliance on platform regulation (Jones, 2018; Stricklin & McBride, 2020). Meanwhile, the underlying technology continues to evolve rapidly, outpacing regulatory efforts (Ferrara et al., 2016; Howard et al., 2018; Shao et al., 2018), emphasizing the critical importance of ongoing research in this domain.

### 2.2 Social bots with a political agenda

Although most social bots are automated to carry out simple, repetitive tasks (Ferrara et al., 2016), political bots are used for malicious purposes (S. Woolley, 2020). They can monitor user traffic while following a circadian rhythm to mimic real users (Cai et al., 2022). The more they act human-like, the more likely they receive engagement from humans (Wischnewski et al., 2022). They can act alone or coordinated as botnets, amplifying or diminishing targeted viewpoints (Shao et al., 2018; Stella et al., 2018). While their interference in political campaigns and electoral processes has been evidenced around the world (Boichak et al., 2021; Bruno et al., 2022; Castillo et al., 2019; Fernquist et al., 2018; Howard et al., 2018; Santini et al., 2021; Stella et al., 2018; Uyheng et al., 2021), they were also found to be actively polarizing across a range of social issues, such as immigration (Nonnecke et al., 2022), climate change (Daume et al., 2023), the vaccine debates ( Yuan et al., 2019), the recent Covid-19 outbreak (Antenore et al., 2023; Chang & Ferrara, 2022), and the war in Ukraine (Zhao et al., 2023).

A large part of the previous work on algorithmic manipulation focused their efforts on bot detection methods, which evolved significantly since the initial attempts in 2010 (Ratkiewicz et al., 2011; Yardi et al., 2010). Over time, researchers have explored different detection strategies, grouped under three main categories: the crowd-based approach, the network-based approach, and the future-based approach (Ferrara et al., 2016; Varol et al., 2017). Crowdsourcing approaches rely on human judgment to identify bots but this approach comes with potential human annotation biases, in addition to issues with scalability and user privacy (Ferrara et al., 2016; H. Y. Yan et al., 2021). Network-based detection involves analyzing the structure of social networks and detecting automated agents through their connection patterns (Cao et al., 2012). This approach assumes bots establish close connections with each other, but the literature suggests this approach may fail to capture automated agents integrated within genuine online communities (Bessi & Ferrara, 2016). Feature-based detection systems leverage a wide array of account characteristics to assess automation through supervised machine learning (Davis et al., 2016; Kudugunta & Ferrara, 2018), however, this approach necessitates continuous refinements to remain effective due to the ever-evolving nature of bots (Subrahmanian et al., 2016; Yang et al., 2019). Recent work comparing bot detection strategies reveal that different approaches can produce markedly different results, further compounding the issue (Beatson et al., 2023; Martini et al., 2021). Overall, bot detection remains an ongoing challenge, requiring constant adaptation from the research community.

Another major theme in the previous literature was bot influence over discussion networks. Bots were found to actively disseminate misinformation during elections (Bessi & Ferrara, 2016; Ratkiewicz et al., 2011), amplifying the reach of low-credibility content such as fake news (Shao et al., 2018). Studies show that bots spread polarizing content, fostering fear, hate and violence (Luceri et al., 2019; Stella et al., 2018). They are centrally positioned in the discussion networks to influence ongoing discourse through retweets, especially during divisive events (Bessi & Ferrara, 2016; Ross et al., 2019; Schuchard et al., 2019). Even weakly connected bots can impact discussions by engaging with influential users (Stella et al., 2018), alter network sentiments (Hagen et al., 2022), sway network dynamics via their followers (Keijzer et al., 2021), and indirectly shape discussions by influencing search engines and recommender systems (Pescetelli et al., 2022).

However, it should be noted that the effectiveness of bot manipulation is contested. Bastos and Mercea (2019) suggests that their influence over the entire discussion may be overstated, while a more recent study shows bots occupy less central positions within the discussion networks, compared to verified influential accounts (González-Bailón et al., 2021). There is evidence to suggest that, while bots do amplify certain divisive narratives, they fail at diffusion, as most of their content fails to reach new users beyond their existing audience (Boichak et al., 2021). Users that are already predisposed to extreme views are most likely to encounter bot-generated content, suggesting minimal shifts in overall public opinion and sentiment (Bail et al., 2020).

At the individual level, research demonstrates that individuals exhibit perceptual biases concerning the prevalence and impact of bots, exaggerating their presence and others' susceptibility to their influence (Yan et al., 2023). Partisan-motivated reasoning also seems to play an important role, as users are more likely to engage with human-like bot accounts when they share the same partisanship (Wischnewski et al., 2022), and less likely to examine the account when they share the same stance (Ngo et al., 2023). Despite minimal direct interactions between bots and users on Twitter, a study of over 4,000 users highlighted bots' substantial influence on opinions through indirect exposure, especially on contentious topics (Aldayel et al., 2022). This underscores a general lack of proficiency among users in distinguishing bots from humans (Beatson et al., 2023), often leading to misidentification (Kenny et al., 2024).

Although there is ample evidence regarding bot activity on Twitter, debates about the impact of these malicious activities are yet to be settled (Duan et al., 2022; González-Bailón et al., 2021; Keijzer et al., 2021). While some researchers have been warning about their increasing sophistication (Boshmaf et al., 2011), and recent evidence suggests that there is the potential for political bots to get more advanced (Cresci, 2020; Luceri et al., 2019), others have found that the majority of the currently available commercial services and tools only provide rather simplistic and repetitive automation (Assenmacher et al., 2020). Most of the Twitter bots in a recent study were found to be "spammers", with no advanced capabilities and limited intelligence (Assenmacher et al., 2020). However, this does not necessarily indicate that their limited capacities could cause no harm.

One possible impact of political bots could be on the overall communication process as a whole; a potential that has largely been underexplored in previous work.

### 2.3 Online Political Communication as a Dynamical Process

Political bots, like all other bots, are automated scripts that follow a predefined path and act within a defined rule set. In a way, they epitomize the concept of predictability by design. And at first glance, the deterministic nature of political bots may suggest a potential to diminish uncertainty in online political communication, making it more predictable through repetitive actions governed by the parameters preset by their developers. Bot activity, by its preprogrammed, repetitive essence, should logically reduce the complexity of online interactions. However, despite their inherently predictable behaviours, the presence of political bots in social media may also be introducing a level of complexity and unpredictability into the online political discourse.

Claude Shannon's foundational framework (1948) in information theory introduces a principle that may be counterintuitive at first: the value of information within a message is inversely proportional to its predictability. In other words, the value of information is quantified not directly by the data transmitted, but in the surprise it delivers to the recipient. The more surprise, the more uncertainty got reduced, the more information got communicated. To illustrate, suppose that we are transmitting sequences that include only two characters 'A' and 'B'. The message 'ABABABA…' is more predictable than a rather random series of 'ABBBAABB…'. Repetitive and predictable structures conveys less information. If it is more difficult to predict the next letter, there is more uncertainty.

In this sense, information theory has long been applied to quantify predictable patterns in the analysis of dynamical systems. This makes it possible to quantify how much information and uncertainty are created, and passed along at each step of a dynamical process. The methods derived have found applications in a variety of fields, such as physics (Crutchfield & Feldman, 2003), biology (Frank, 2012), neuroscience (Gallistel & Matzel, 2013), psychology (Heath, 2014) and communication (Hilbert & Darmon, 2020a). In this framework, the unpredictable aspect of a process is quantified with the traditional concept of uncertainty (i.e. entropy), as proposed by Shannon (1948). The amount of uncertainty that can be reduced by identifying a pattern is quantified by the predictable information. And the 'machinery' that creates such pattern can be characterized by predictive complexity, i.e. the complexity required to produce the predictable pattern (Crutchfield & Feldman, 2003).

Although this approach has relevance to various aspects of communication science, only a few scholars have explicitly utilized this framework to analyze the multi-level dynamics of online communication (Waldherr et al., 2021). Consequently, despite the fact that Shannon's original paper is literally called "A Mathematical Theory of Communication" (1948), the application of information theory to communication as a dynamical system has been limited. In this study, we consider online political communication as an example of a dynamic process.

### 3. Research Questions

On one hand, the simplicity of political bot behaviors, as evidenced in the previous work, might suggest a decrease in the complexity of communication dynamics. Most bots engage in simple amplification tasks, such as liking and resharing existing content to simulate social contact (Assenmacher et al., 2020; Schuchard et al., 2019; Yang et al., 2019). Considering that they can efficiently perform such repetitive tasks in bulk and at a much higher frequency than human users (Grimme et al., 2017; Schuchard et al., 2019), it could be argued that political bots are likely to make the overall communication dynamics more predictable, thereby decreasing the complexity and uncertainty inherent in the communicative process.

Conversely, the evidence regarding their capability to trigger emotional contagion around negative topics (Shi et al., 2020), or in spreading polarizing, sensational content (Chang & Ferrara, 2022; Shao et al., 2018; Yan et al., 2021), suggests a compelling counterargument. Such activities by political bots may be introducing significant levels of complexity and unpredictability into online political communication, seemingly increasing the system entropy.

Viewed through the lens of information theory, the level of complexity in online political communication can be determined by the frequency of recurring patterns that emerge during the dynamic process (Crutchfield & Feldman, 2003), while uncertainty is the measure of average ambiguity that is dependent on the complexity of the process (1948). It has been previously shown that social bots can have an impact on dynamic processes. In a study on the communicative turns of edits on Wikipedia, Hilbert and Darmon (2020) reported that editor bots made the structural dynamics of the communicative process more predictable, more complex and, at the same time, more uncertain. They found that bots resolved much of the previous uncertainty (i.e. uncertainty related to vandalism or missing signatures). At the same time, bots fine-grained the dynamic, both in space (variety) and time (patterns), which created new complexity and induced new uncertainties in the process.

But editor bots operate in an entirely different context than a Twitter bot with a political agenda, and as a result, their respective impact may vary significantly. In this work, we adopt the same approach and expand on this analytical framework to examine the role of political bot activity in online political discussions. Unlike previous studies, we delve into the underlying complexity and uncertainty these digital agents introduce into the online political discussion. By quantifying the information dynamics of online political interactions, we measure the complexity and uncertainty involved in the process and seek to answer below questions:

RQ1: How is the presence of bots associated with the level of complexity in online political communication dynamics?

RQ2: How is the presence of bots associated with the level of remaining uncertainty in online political communication dynamics?

## 4. Data & Methods

We perform a multiple regression analysis with the two information theoretical measures as our dependent variables (quantifying how complex and how uncertain is the communication dynamics), and with bot-level as our explanatory independent variable, controlled by some potential confounders, like word count, character count, word complexity and time variance.

The data was collected real-time during the first democratic presidential debate in 2019 by leveraging Twitter's search API with a keyword approach. The dataset provides roughly 395K tweets that include the keyword '#demdebate'. To represent the resulting stream of tweets quantitatively, we semantically analyze them using IBM Watson Developer Cloud's natural language understanding service, for a representation of each tweet in terms of the five primary emotions conveyed, mainly anger, fear, sadness, joy and disgust (Ekman et al., 1969). The IBM Watson API is a natural language processing tool that employs machine learning algorithms to extract semantic information from text, including concepts, emotions and sentiments. Rather than relying on traditional methods, the application utilizes a Recurrent Neural Network to dynamically capture and classify sentiments. The API has been trained on a diverse set of sources, including tweets, and has been used in numerous research projects (Hilbert et al., 2018; W. Zhang et al., 2020).

Since we work with traditional information theory, we convert the raw scores into categorical variables (otherwise we would have to work with differential/continuous entropy). We assign each tweet into one of four categorical bins based on its emotional score from IBM Watson, which ranged from 0 to 1. This leaves us with five temporal sequences of roughly 395K tweets, one for each emotion (i.e. anger, fear, sadness, joy and disgust). We want to analyze statistical results, so we need more than five sequences, but our sequences also need to be sufficiently long, because our nonlinear information theoretic measures converge rather slowly. We find a statistically robust sampling solution by creating sequences of 3,000 consecutive tweets for each emotion, ending up with 650 temporal sequences in total. Despite sampling motivations, it is methodologically useful that each sequence is equally long, as longer/shorter sequences will increase/decrease the likelihood of more structural diversity and more/less uncertainty.

To ensure statistical robustness and gain a nuanced understanding of the data distribution, we employ three different binning strategies in categorizing the emotional scores of each tweet in each sequence. This was deemed necessary to account for variations in the distribution of emotion scores, as how we bin the data into categorical variables might affect our analysis. The first binning strategy involves categorizing each raw value in the sequence by dividing it into quartiles, which creates four categories of approximately equal sizes, where the first category includes values that are less than the 25th percentile and the last category includes values greater than the 75th percentile. The second binning strategy involves sorting and ranking the values in each sequence, and creating uniform, equal-sized groups for each of the four categories. The groups are then de-indexed, meaning they return to their original position in the sequence. The third binning strategy follows an exponential approach, in which the width of each one of the four categories is determined by an exponential function. Each subsequent category is wider than the previous one, creating an increasing distribution of the data across categories.

**Dependent variables**

We calculate two complementary measures adopted from information theory that quantify the dynamics of complexity and uncertainty for each of the 650 temporal sequences. We use an empirically validated algorithm to compute the two measures from our data, namely the Causal State Splitting Reconstruction (CSSR) algorithm (Darmon, 2020), which reliably infers the statistical structure of a given dynamical process (Darmon, 2015; Shalizi & Shalizi, 2004). It produces three outputs: predictive information $E,$ predictive complexity $C$ and remaining uncertainty $h,$ though in this study, we are mostly concerned with complexity and uncertainty. All variables are measured in bits, which gives the optimal, or average minimum number of yes-or-no questions that an observer would have to ask to completely reconstruct a system. The below Venn diagram, based on Crutchfield et al. (2009), portrays the relationship between these three complementary measures along temporal sequences of past and future (Figure 1a).

Within this formalism, predictive information is the common information between past and future states; that is, the information preserved by the dynamical system over time. The predictive complexity (C) includes the information from the past that is necessary to predict the predictable component of the future, and the remaining uncertainty (h) is the amount of information about the future that cannot be derived from the past behavior (possibly random, possibly predictable on the basis of outside information). To derive the statistical measures of character sequences in our data, we use sliding windows of three characters as depicted in Figure 1.b, and measure the occurrence of unique character subsequences. These measures are calculated with the Python implementation of CSSR by Darmon (2015).

**Figure 1**

*Diagram depicting the relationship between predictive information E, predictive complexity C and remaining uncertainty*

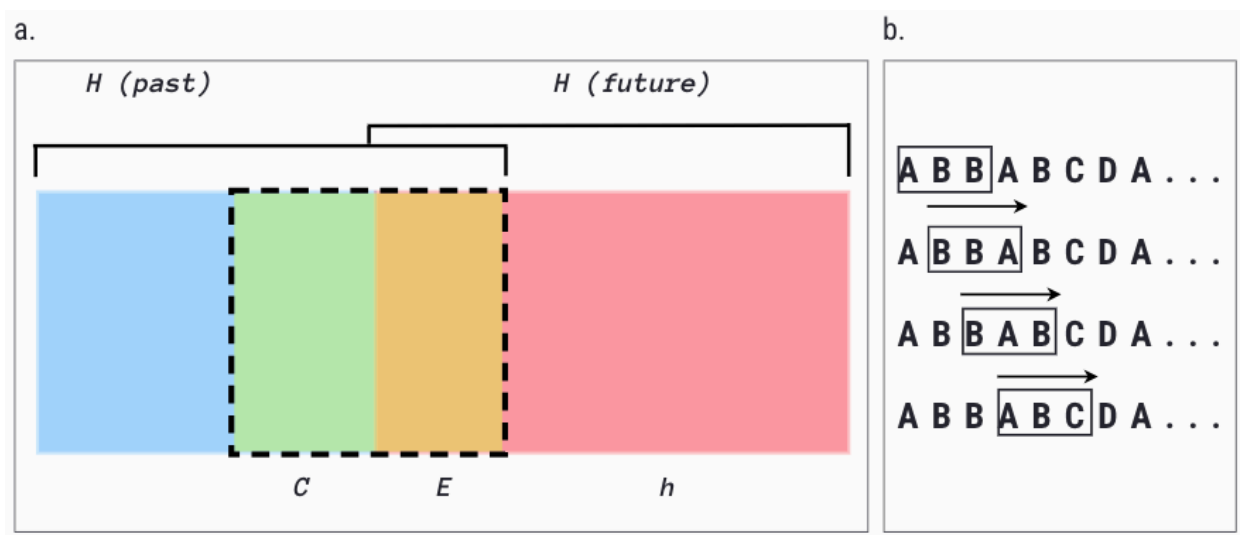

*Note.* a) This Venn diagram portrays the relationship between predictable information *E,* predictive complexity *C* and remaining uncertainty *h*. The amount of information communicated

from the past to the future through repeating patterns is quantified with $E$, the predictable information. The amount of information required to produce E is quantified by the complexity of the pattern, $C$. The remaining uncertainty, $h$, is the entropy rate, which quantified the amount of uncertainty of the future dynamic that cannot be predicted by identifying all possible patterns contained in the past. b) On the right, the illustration depicts the sliding-window approach we employed to measure the frequency statistics from character sequences. The observed frequency of these subsequences, which are essentially the fundamental components of the dynamic process, provide us with the statistics that outline the features of the underlying temporal structure, with a focus on how information is being preserved, created, and destroyed at each moment of the dynamic.

### *Predictive complexity (C)*

Predictive complexity, expressed with $C$, is also known as statistical complexity (Crutchfield & Young, 1989) and approximates Kolmogorov complexity (Crutchfield, 2011). It quantifies the minimum amount of information that can predict the amount of structure from the past that is useful to predict the future (or "predictable information," $E$ in Figure 1a). Essentially, it is measured as the amount of bits needed, to optimally predict the future of the process from its past (Crutchfield et al., 2009). For instance, $C$ would be zero for an entirely random process. The higher the required number of bits to predict the future, the more complex the process is. Complexity is therefore defined as the information required to describe a process "between order and chaos" (Crutchfield, 2011, p. 20). It corresponds to the creation or preservation of information or structure in the signal. The mathematical beauty of the measures is that it is shown that the predictive complexity is a sufficient statistic, with the minimal size representation of the structure, but with maximal predictability of what can be predicted about the process (Crutchfield & Young, 1989; Shalizi & Crutchfield, 2001). C is a practical approximation of the Kolmogorov complexity of a dynamic, in the sense that it measures the size of the smallest model with maximal predictive power for the time series (Crutchfield, 2011).

### *Remaining uncertainty (h)*

The remaining uncertainty, expressed with the traditional measure of the entropy rate *h,* is also known as the entropy rate of the process and is a fundamental measure of dynamical systems (Kolmogorov, 1959; Sinai, 1959). It quantifies the amount of uncertainty per symbol about the future that remains in a process, given all the information the previous state of the process can tell us about the future. It is essentially the rate of conditional entropy $H$ calculated per symbol, measuring the uncertainty involved in predicting the next symbol based on the previous one (Shannon, 1948). The resulting entropy rate measures the uncertainty of the next turn conditioned on the previous turns (Cover & Joy, 1991). In short, the entropy rate "gives the source's intrinsic randomness, discounting correlations that occur over any length scale" (Crutchfield, 2011). Naturally, the higher the entropy rate, or remaining uncertainty, the higher the probability of a prediction error, which makes the future of the process less predictable (Hilbert et al., 2018).

### Independent variables

The main independent variable in our study is botness, or "bot level", the extent to which an account is estimated to be a bot. To assess the automation level involved with each account, we used a publicly available bot detection service frequently used in previous work, the Botometer API (Varol et al., 2017; Yang et al., 2022). Using machine learning algorithms, Botometer extracts over 1000 predictive features that identify numerous suspicious behaviors by characterizing an account's profile, social network, friends, temporal activity patterns, language and sentiments (Davis et al., 2016; Yang et al., 2022). It then provides a classification score on a normalized scale, indicating the probability that a Twitter account is likely to be a bot. Scores closer to 1 indicate a higher probability of the account being a bot, while those closer to 0 suggest that the account is controlled by a human. We averaged the resulting scores for each tweet per sequence. The variable bot-level measures the percentage of *botness* within each of the 650 temporal sequences made up of 3000 consecutive tweets (M = 0.045, SD = 0.008, min = 0.024 max = 0.081). The continuity of this measure captures not only uncertainty about the conclusion that an account is a bot account, but also the fact that accounts vary in their botness, as in the case of accounts that tweet a mix of human and automated content.

We also control for several potential confounding variables to account for the characteristics of tweets. One such confounding factor that could affect the complexity and uncertainty of a communication dynamic is the number of words used. This measure could contribute to complexity, for longer tweets with more words could contain multiple clauses with complex sentence structures, potentially increasing the level of information contained. The variable "word count" is measured by calculating the number of words used in each tweet, and averaging the total count per each temporal sequence (M = 21.219, SD = 0.456, min = 19.879, max = 24.18).

Another factor we consider is the expression complexity per word. Longer and more complicated words used in tweets can indicate a more complex sentence structure, while shorter or simpler words can drastically reduce the amount of information conveyed. The variable "word complexity" is created by dividing the number of characters in each tweet by the number of words per tweet, averaged per sequence (M = 6.497 SD = 0.127, min = 6.318, max = 6.778).

Finally, we control for time variation between tweets, for a high time-variance could make it more difficult to predict how future states will be related to the past, which can affect the overall uncertainty and predictability. We therefore measure the variable "time-variance" to control for the variation of time between the first and last tweet across each temporal sequence (M = 14.4, SD = 117.5, min = 0.477, max = 1338.7).

## 5. Results

We begin our analysis with preliminary diagnostic tests on our three binning strategies to assess their statistical assumptions. The first strategy, which involves categorizing raw values in the sequence by dividing them into quartiles, outperforms the other two in terms of normality, outliers, heteroscedasticity, and low collinearity. Although the assumption of linearity is always a concern when working with social media data, our classical statistical assumption tests reveal that the first strategy appears to be more promising in every aspect. Therefore, we proceed with our analysis using this first approach.

Moving onto analysis, we first assess the relationship between bot presence and the predictive complexity, reflecting how much information from the past is required to foresee the future state. Overall, the regression model is significant (F(5, 641) = 38.73, p < 0.001, $R^2 = 0.232$, Figure 2a). Our results suggest a significantly positive relationship between the presence of bots and predictive complexity ( $\beta = 0.151, p < 0.001$, Figure 2). Notably, higher levels of bot presence correspond with higher levels of complexity. Put simply, online political communication sequences with higher bot levels require more past information to predict future behavior. The presence of bots creates more complex patterns.

**Figure 2**

*Standardized regression coefficient estimates*

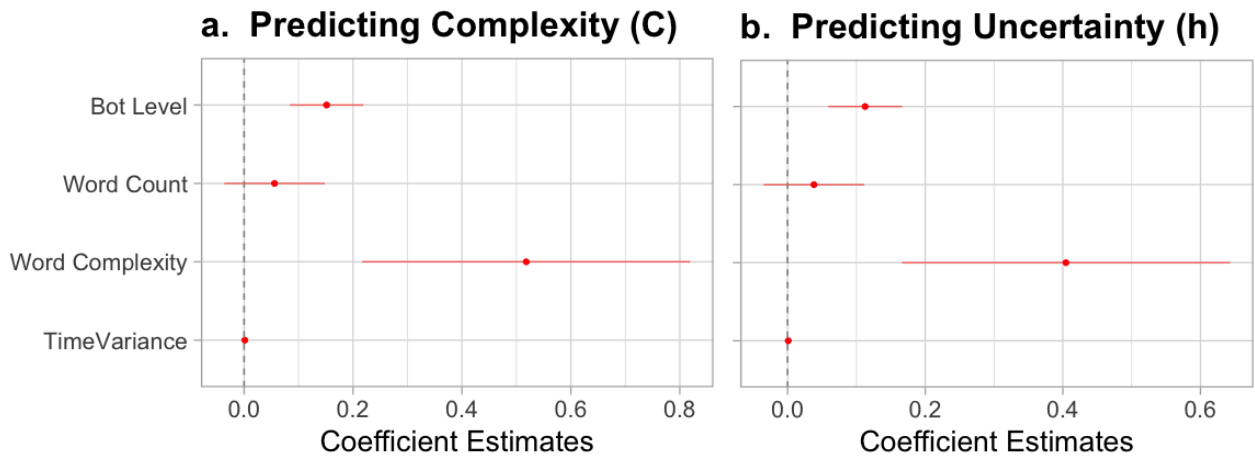

a. **Predicting Complexity (C)**     b. **Predicting Uncertainty (h)**

*Note.* Standardized regression coefficients for a) predicting complexity and b) remaining uncertainty bound with 95% confidence intervals.

Additionally, we find that both the complexity of the language used in tweets, or word complexity, ($\beta = 0.517, p < 0.001$) and time variance ($\beta = 0.003, p < 0.001$) positively and significantly contribute to predictive complexity of the ongoing discussion. Though the contribution of time variance is comparatively much smaller. Also, results show a positive relationship between word count and predictive complexity but his relationship lacked significance ($\beta = 0.055, p = 0.23$).

Turning to the remaining uncertainty involved in the online political discourse, our second model is also significant ($F(5, 641) = 35.08$, $p < 0.001, R^2 = 0.214$, Figure 2b). Notably, higher bot levels correspond with increased levels of uncertainty ($\beta = 0.113, p < 0.001$). Controlling for all the other predictors we include in the model, higher bot levels significantly correlate with higher remaining uncertainty: the higher the bot levels in sequences of tweets, the larger the component of future behavior that cannot be predicted from past states. As for the control variables, we find a positive relationship between uncertainty and word complexity ($\beta = 0.404$, $p < 0.001$), and a small effect from time variance ($\beta = 0.001$, $p < 0.001$). But the positive relationship between word count and uncertainty lacked significance ($\beta = 0.04$, $p = 0.3$). [Figure 2 here]

These findings illustrate a significant correlation between bot levels in the temporal sequences and communication dynamics. As bot presence in sequences of tweets increases, so does the challenge predicting the semantic trajectory of future engagements based on past data. Our results align with the argument that political bots introduce both structural uncertainty and complexity into online discussions revolving around political issues, altering the stability of online political discourse.

## 6. Discussion

This study sheds light on the the structural changes in online political discussion dynamics attributable to bot activity, by analyzing the relationship between bot presence, remaining uncertainty, and predictive complexity. Our results illuminate the impact of political bots on macro-level dynamics of online political discourse, demonstrating that increased bot activity is associated with higher levels of structural complexity and uncertainty.

By applying Claude Shannon's information theory (1948) within a dynamical systems framework, we offer a novel lens through which to view online political communication. This approach reveals that political bots not only contribute to the noise within the digital public sphere but also complicate the process of political discourse by making it less predictable and more chaotic. Contrary to their intended function of simplifying communication through automation and repetition (Franklin & Graesser, 1997; Gorwa & Guilbeault, 2020), politically motivated bots significantly contribute both more complexity and uncertainty within online discourse. This increased structural complexity and uncertainty can have profound implications for democratic engagement, potentially hindering users' ability to engage in meaningful discussions and access important information during critical events.

Our results suggest that higher bot activity is associated with a larger component of future behavior that cannot be predicted from the past states, indicating more uncertainty in the process. This increase in structural uncertainty due to higher bot activity suggests a fundamental disruption in the predictability of communication patterns. In practical terms, the usual cues and patterns that individuals rely on to interpret messages and intentions online could become obscured. The presence of bots introduces noise and unpredictability, which could affect the ability of users to engage in reasoned debate or to follow the progression of events online. This may not only affect the immediate understanding of specific conversations or the perceptions of certain critical political events but could also erode trust in the digital communication environment as a whole, as users increasingly become uncertain about which interactions are genuine and which are manipulated to control public opinion (Kenny et al., 2024; Ngo et al., 2023; Wischnewski et al., 2022).

Higher bot activity also requires more past information to predict the same amount of future behavior, indicating a greater degree of complexity in the process. This complexity is not necessarily a challenge in terms of the volume or density of information, but rather it shows a substantive change in the nature of the discourse itself. This additional layer of complexity could make it more difficult for both individuals and algorithms (such as those powering recommendation systems) to filter and understand the essential elements of the ongoing conversations. Earlier work shows that bots can substantially shape discussions by influencing search engines and recommender systems (Pescetelli et al., 2022), their impact on communication dynamics may exacerbate these effects. Moreover, bots' potential ability to skew the semantic direction of the discourse may distort the public's perception of consensus or controversy on critical issues, further polarizing audiences. The outcome is a further distorted representation of public opinion, where artificially amplified themes and emotions gain undue prominence, disrupting discussions and potentially misleading observers about the true nature of public sentiment around controversial topics.

Our analysis also reveals that the complexity of the language used in tweets (word complexity) and the intervals between posts (time variance), also significantly influence both the complexity and uncertainty of online political discussions, though the influence of time variance is rather weak. Conversely, the volume of words used (simple word count) does not show significance. These results align with and support previous research emphasizing the value of quality engagement in sustaining the vitality and success of online communities and digital platforms, over frequency or quantity of engagement (Butler, 2001; Cunha et al., 2019; Kollock & Smith, 1996).

The present study makes several important contributions. Our study's focus on the macro-level dynamics of an online political discussion provides an important expansion of the current work on social bots. Previous research has largely focused on the prevalence, reach, detection, and influence of social bots, such as their role in spreading misinformation (Bessi & Ferrara, 2016; Bradshaw et al., 2020), interfering in political campaigns (Bastos & Mercea, 2019; Howard et al., 2018), and amplifying polarizing viewpoints (Nonnecke et al., 2022; Zhao et al., 2023). While these studies provide crucial insights, they often overlook the broader structural implications of these automated entities on the digital eco-system. By analyzing how bots contribute to the structural complexity and uncertainty of online political discourse, our work offers a holistic view of their influence, underscoring how bots fundamentally alter the dynamics of online political communication.

Our study also contributes to the growing literature on information theory in understanding communication processes. By extending the dynamical systems framework to political bots and using information-theoretic measures to quantify the complexity and uncertainty of online political communication, we expand on the previous work and provide a novel approach to studying the impact of political bots. This approach could be further extended to other domains, such as crisis communication and health communication, and could provide insights into the macro-level dynamics of online communication that previously went unexplored.

Furthermore, this study applies the Causal State Splitting Reconstruction (CSSR) algorithm (Darmon, 2020) to analyze online political communication. This algorithm has previously been used in other fields to analyze dynamical systems, and our study demonstrates its potential for use in the analysis of online communication on social media platforms.

## 7. Limitations

First and foremost, it is important to note that we cannot establish a causal relationship among the variables we considered in this work, and there is a chance that a different variable, beyond those we control for in this study, plays a causal role. However, the direction of the bivariate relationship implies causality on theoretical grounds. It is more likely that the presence of political bots inflicts structural changes in the dynamics of online political communication, than that the complexity or uncertainty involved in tweet sequences increases bot presence.

Furthermore, our study only provides a snapshot of an online political discussion, and the Twitter API we used in data collection only provides 1% of the total traffic on Twitter. This limited sample may not fully capture the dynamics of online political communication, and findings may not generalize across different platforms, events, or contexts. Each platform's unique dynamics and algorithms can influence bot and user behavior differently. Further research is needed to understand the broader effects of bot activities on various social and political discussions.

Another limitation is using Botomer to classify the bot levels of each account in our data set. Despite its popularity, Botometer is prone to errors and it is known to have several limitations, as is the case with many other supervised machine learning algorithms (Martini et al., 2021). For instance, botometer may sometimes struggle dealing with new accounts that heavily differ from those in its training dataset, or those that use different languages other than English (Yang et al., 2022). We use continuous measures in our analysis to represent the bot level of each account with the raw scores that show the complete automation probability assessed by the Botometer, but the possibility of bias may remain. The evolving nature of bot programming means some sophisticated bots might have been missclassified or overlooked. Further work could explore different bot detection techniques to contrast and validate our results.

## 8. Conclusion

This study provides evidence for the macro-level impact of political bots using a dynamical systems framework. Our findings reveal that political bots structurally influence online political discussions, increasing complexity and uncertainty. These results highlights the importance of studying communication processes as complex and dynamic systems, and the need for further research into the macro-level influence of bots. While bots continue to be increasingly prevalent in the digital sphere, it becomes increasingly important to understand their potential impact on online political discourse and develop effective strategies to mitigate their malicious activities. By adopting a complex systems perspective on bots and online political communication, we hope to contribute to a more nuanced understanding of the interplay between technology and society in the digital age.